\documentclass[11pt]{article}

\usepackage{times}
\usepackage{setspace}
\usepackage{bbm}
\usepackage{graphicx,verbatim,array,multicol,fontenc,subfigure,pifont}
\usepackage{psfrag, fancybox, subfigure}
\usepackage{amsmath, amssymb, epsfig, color,mathrsfs,latexsym}
\usepackage{soul}

\usepackage[T1]{fontenc}
\usepackage{graphics}
\usepackage{latexsym}
\usepackage{rotating}
\usepackage{amsmath}
\usepackage{color}
\usepackage[latin1]{inputenc}
\usepackage{pstricks,pst-node,pst-text,pst-3d}
\usepackage{lscape}

\usepackage{mathpazo}
\usepackage{fullpage}
\usepackage{psfrag}
\usepackage{pifont}

\usepackage{amssymb}
\usepackage{amsthm}
\usepackage{indentfirst}
\usepackage{slidesec}
\usepackage{graphicx}
\usepackage{epsfig}
\usepackage{fancybox}
\usepackage{color}
\usepackage{amssymb,amsmath}
\usepackage{epsfig, graphicx}
\usepackage{epsfig}
\usepackage{mathrsfs}
\usepackage{bbm}

\usepackage{subfigure}
\usepackage{booktabs}
\usepackage{setspace}
\definecolor{green}{RGB}{000,150,100}
\definecolor{purple}{RGB}{220,000,250}
\definecolor{turquoise}{RGB}{0,206,209}
\definecolor{gray}{RGB}{150,150,150}

\def\beqn{\begin{eqnarray}}
\def\eeqn{\end{eqnarray}}
\def \tpf{\text{TPF}}
\def \ppv{\text{PPV}}
\def\npv{\text{NPV}}
\def \ap{\text{AP}}
\def \auc{\text{AUC}}
\def \ppvtilde{\widetilde{\ppv}}
\def\ppvhat{\widehat{\ppv}}
\def\tpfhat{\widehat{\tpf}}
\def\tpfhat{\widehat{\tpf}}

\def \what{\widehat{w}}
\def\Gcal{\mathcal G}
\def\Gcalhat{\widehat{\Gcal}}

\def \Rcal{\mathcal{R}}

\begin{document}

\title{A Threshold-free Prospective Prediction Accuracy Measure for Censored Time to Event Data}

\author{YAN YUAN$^{\ast \dag}$\\
{School of Public Health, University of Alberta, Edmonton, AB T6G1C9, Canada}\\
\and QIAN M. ZHOU$^\dag$\\
Department of Statistics and Actuarial Science\\
Simon Fraser University, Burnaby, B.C.  V5A1S6, Canada\\
Department of Mathematics and Statistics\\
Mississippi State University, Starkville, Mississippi 39762, USA\\
\and BINGYING LI\\
Department of Statistics and Actuarial Science\\
Simon Fraser University, Burnaby, B.C.  V5A1S6, Canada \\
\and HENGRUI CAI\\
School of Public Health, University of Alberta, Edmonton, AB T6G1C9, Canada\\
\and ERIC J. CHOW\\
Fred Hutchinson Cancer Research Center\\
Seattle Children's Hospital, University of Washington, Seattle, Washington, USA\\
\and GREGORY T. ARMSTRONG \\
Department of Epidemiology and Cancer Control, Division of Neuro-Oncology\\
St. Jude Children's Research Hospital, Memphis, TN 38105, USA\\
 }

\markboth%
{Y. Yuan, Q. Zhou and others}
{Time-dependent AP}

\maketitle

\footnotetext{ To whom correspondence should be addressed. $\dag$ Dr. Yuan and Dr. Zhou contributed equally to this work.}
\newpage

\begin{abstract}
Prediction performance of a risk scoring system needs to be carefully assessed before its adoption in clinical practice. Clinical preventive care often uses risk scores to \textit{screen} asymptomatic population. The primary clinical interest is to predict the risk of having an event by a pre-specified \textit{future} time $t_0$. Prospective accuracy measures such as positive predictive values have been recommended for evaluating the predictive performance. However, for commonly used continuous or ordinal risk score systems, these measures require a subjective cut-off threshold value that dichotomizes the risk scores. The need for a cut-off value created barriers for practitioners and researchers. In this paper, we propose a threshold-free summary index of positive predictive values that accommodates time-dependent event status. We develop a nonparametric estimator and provide an inference procedure for comparing this summary measure between competing risk scores for censored time to event data. We conduct a simulation study to examine the finite-sample performance of the proposed estimation and inference procedures. Lastly, we illustrate the use of this measure on a real data example, comparing two risk score systems for predicting heart failure in childhood cancer survivors.
\end{abstract}

Censored event time; Positive predictive value; Precision-recall curve; Risk prediction; Screening; Time-dependent prediction accuracy

\maketitle
\footnotetext[2]{Dr. Yuan and Dr. Zhou contributed equally to this work.}

\section{Introduction}
\label{sec:intro}
Clinical medicine is facing a paradigm shift from current diagnosis and treatment practices to prevention through earlier intervention based on risk prediction \cite{wright2014conceptual}. Diagnosis and treatment approaches help individual patients seek relief from their symptoms. However, evidence is mounting that health interventions may be more effective in improving long-term health outcomes when they target asymptomatic individuals who are predicted to be at high risk for the condition of interest \cite{espeland2007reduction,james2010early}. The condition of interest typically has the following characteristics: 1) its seriousness may result in a high risk of mortality or significantly affect the quality of life; 2) early detection/intervention can make a difference in disease prognosis; and importantly but subtly 3) its event rate is low. A prevention approach to medicine relies on the development of risk scores to stratify individuals into different risk groups. Early intervention strategies are typically recommended to subjects who are in the high-risk group.

In the prevention paradigm, the use of risk scores as population \textit{screening} tools is increasingly advocated in clinical practices, e.g. \cite{goff20142013}. For example, one systematic review identified forty-six algorithms that predict the risk of type 2 diabetes \cite{buijsse2011risk}. Another study established several risk score systems to predict congestive heart failure for childhood cancer survivors who are at an elevated risk due to treatment toxicity \cite{chow2015individual}. One of the defining characteristics of screening is a low event rate in the targeted asymptomatic population. Taking the aforementioned two diseases as an example, the crude prevalence of undiagnosed type 2 diabetes, a common disease, was low at 3.5\% in 1987 and 5.7\% in 1992 \cite{lindstrom2003diabetes}, while the cumulative event rate of congestive heart failure by 35 years post childhood cancer diagnosis was 4.7\% \cite{chow2015individual}. The event rate is much lower for other notable conditions such as cancer, multiple sclerosis, AIDS, dementia, COPD and perinatal conditions. A low event rate and a focus on prevention necessitate the development of \textit{screening} tools such as risk scores.

Before a risk scoring system is adopted for clinical screening, evaluation of its predictive accuracy is critical. The most popular accuracy metric used in the clinical literature is the area under the receiver operating characteristic (ROC) curve (AUC). The AUC is a summary index of two accuracy metrics -  sensitivity and specificity -  which are both retrospective metrics. Thus, the AUC does not reflect the prospective predictive accuracy of risk score systems. Indeed, one influential article criticized these retrospective metrics as being of little use for clinicians because clinical interest almost always focuses on prediction \cite{grimes2002uses}. In contrast, a prospective accuracy measure, such as positive predictive value (PPV), can prospectively answer the question: "Can risk scores be trusted?" Unfortunately, a risk score with high sensitivity and specificity, and thus a high AUC, can have poor PPV when applied to low-prevalence populations. This limitation is often overlooked by clinicians and biomedical researchers. Despite its popularity, studies confirm that the AUC is insensitive in evaluating risk prediction models. For example, including a marker with a risk ratio of 3.0 showed little improvement on the AUC, while it could shift the predicted 10-year disease risk for an individual patient from 8\% to 24\% \cite{cook2007use}. This range would result in different recommendations on follow-up/intervention strategies.

Compared to the AUC, the PPV provides a more appropriate assessment of the prospective prediction performance of the risk score \cite{moskowitz2004quantifying}, making the PPV a superior metric for risk score systems used as screening tools. The PPV is calculated with data from a prospective cohort, where the risk scores are computed using baseline information and the outcome is followed prospectively. Originally, the PPV was defined for a dichotomous test. Moskowitz and Pepe (2004) extended the definition of PPV for a continuous risk score \cite{moskowitz2004quantifying}. Assuming that the higher the risk score, the greater the individual risk, the PPV is defined as the probability of having the disease when the risk score value is larger than a given cut-off value $z$,

\begin{equation}\label{equ:PPV-ContMarker-BinOutcome}
\ppv(z) = Pr\{D=1\mid Z \geq z\}\quad \hbox{and}\quad \npv(z) = Pr\{D=0\mid Z < z\},
\end{equation}
where $D=1$ indicates the presence of the disease, and $D=0$ indicates the absence of the disease. Zheng et al. (2008) further generalized the definition to accommodate the censored event time outcome \cite{zheng2008time}. Since the PPV is threshold dependent, as seen in (\ref{equ:PPV-ContMarker-BinOutcome}), it is often evaluated at either several fixed specificities or several fixed quantiles of the . Such evaluations allow the comparison across different risk score systems \cite{moskowitz2004quantifying, wald2014area}. The selection of specificities or quantiles can be subjective, and it is possible that different systems could outperform others, depending on the cut-off points selected  \cite{zheng2010semiparametric}.

For the above reasons, a threshold-free summary metric for the PPV is needed to facilitate its clinical usage. Two curves of PPV have been investigated in the literature. Raghavan et al. (1989) and Zheng et al. (2010) considered a curve of PPV versus quantiles of the risk score \cite{zheng2010semiparametric, raghavan1989critical}. However, they did not provide a summary index of the proposed PPV curve. A second curve is called the precision-recall (PR) curve, which was proposed in the information retrieval community \cite{raghavan1989critical,manning1999foundations}, where precision is equivalent to the PPV and recall is equivalent to the sensitivity. The relationship of PR and ROC curves and the area under them has been discussed in Davis and Goadrich \cite{davis2006}. They showed that the PR curve of a risk score system dominates that of another system if its ROC curve is also dominant. However, such a relationship does not exist for the area under these two curves \cite{davis2006, su2015relationship}. Two recent papers illustrated the advantage of using the area under the PR curve over the AUC for predicting low prevalence diseases \cite{yuan2015threshold, ozenne2015precision}. We refer to the summary metric for the area under the PR curve as the average positive predictive value (AP) \cite{yuan2015threshold}. These previous research on the area under the PR curve have only considered binary outcomes. However, for many clinical applications, the outcome is time to event.

We make three contributions in the assessment of risk scoring systems for clinical screening. First, we define a time-dependent AP, $AP_{t_0}$ for censored event time outcomes. We propose a robust nonparametric estimator of $\ap_{t_0}$ without modeling assumptions on the relationship between the risk score and event time. Secondly, we provide a statistical inference procedure to compare the $\ap_{t_0}$ two risks scores regarding . Thirdly, we provide an R package to implement our method. The paper is organized as follows. In Section~\ref{sec:def}, we introduce the definition and interpretation of $AP_{t_0}$. In Section~\ref{sec:estimator}, we present the inference procedures for estimating $AP_{t_0}$ of a single risk score as well as the comparison between two competing risk scores. In Section~\ref{sec:simulation}, we conduct a simulation study to investigate the performance of the proposed estimation and inference procedures in finite samples. In Section~\ref{sec:data-analysis}, we illustrate the proposed metric $AP_{t_0}$ by analyzing two risk score systems with data from the Childhood Cancer Survival Study \cite{robison2009childhood}. We conclude with a discussion and suggestions for future work in Section~\ref{sec:discussion}.

\section{Time-dependent Average Positive Predictive Values}
\label{sec:def}
Consider a continuous risk score $Z$. Let $T$ be the time to the event of interest. Time-dependent PPV and TPF \cite{zheng2008time, heagerty2000time} are defined as
\begin{equation}\label{equ:PPV-ContMarker-EventTime}
\ppv_{t_0}(z) = Pr\{T<t_0\mid Z \geq z\}\quad \hbox{and}\quad \tpf_{t_0}(z) = Pr\{ Z \geq z \mid T<t_0\}.
\end{equation}
 In the above setting, the event status is time-dependent, i.e., $D_{t_0}=I(T<t_0)$, where $I(\cdot)$ is an identity function. Consequently, the PPV and TPF are also functions of $t_0$.

Following \cite{yuan2015threshold}, we define $\ap_{t_0}$, as the area under the time-dependent PR curve \allowbreak $\{(\tpf_{t_0}(z) ,\ppv_{t_0}(z)), z\in \Rcal\}$,
\begin{equation}\label{equ:TimeDependent-AP}
\ap_{t_0} = \int_{\mathcal{R}}  \ppv_{t_0}(z) d \tpf_{t_0}(z).
\end{equation}
Note that the TPF describes the distribution function of $Z$ in ``cases" who experience the event by time $t_0$, i.e. $T<t_0$. It can be shown that $\ap_{t_0} = E_{Z_1}\left\{\ppv_{t_0}(Z_1)\right\}$, where $Z_1$ denotes the risk score in cases.  In the real data example of Section \ref{sec:data-analysis}, we will show that AP is estimated to be 0.114 at $t_0=35$ years for a risk score system. That is, by 35 years post diagnosis, we expect that on average 11.4\% of the subjects with a high risk score (compared to the risk score of a randomly selected case) will experience the event of interest.

In addition, $\ppv_{t_0}(z)$ can be written as $\ppv_{t_0}(z) =  P( Z \geq z \mid T<t_0)P(T < t_0) / P(Z \geq z) = \pi_{t_0}\left\{1-F_1(z)\right\}/\left\{1-F(z)\right\}$, where $F_1(z)=Pr(Z< z \mid T<t_0)=P(Z_1<z)$ is the distribution function of the risk score $Z_1$ for cases, $F(z)=P(Z < z)$ is the distribution function of the risk score $Z$ for the target population, and $\pi_{t_0}=Pr(T < t_0)$ is the event rate by time $t_0$ in the target population. Thus, the AP can be written as
\begin{equation}\label{equ:TimeDepedent-AP-alt}
\ap_{t_0} = \pi_{t_0} \int_{\mathcal R} \frac{1-F_1(z)}{1-F(z)} d F_1(z).
\end{equation}
A perfect risk score system would always assign higher values to cases, individuals with $T<t_0$, compared to those controls, individuals with $T\geq t_0$, i.e. $P(Z \geq Z_1\mid T \geq t_0) = 0$. This leads to $\ap_{t_0}=1$ from equation (\ref{equ:TimeDepedent-AP-alt}). A non-informative risk score system would randomly assign risk scores to both cases and controls. i.e., $P(Z \geq z\mid T\geq t_0) = P(Z\geq z \mid T<t_0)$ for each $z$, which leads to $\ap_{t_0}=\pi_{t_0}$. Thus, the theoretical range of $AP_{t_0}$ is $[\pi_{t_0},1]$.

\section{Estimating and Comparing $\ap_{t_0}$} \label{sec:estimator}
\subsection{Nonparametric Estimator of $\ap_{t_0}$ for a single risk score}\label{sec:single}
Often, the event times of some subjects are censored due to the end of the study or loss to follow up. Due to censoring, one can only observe $X=\min\{T,C\}$ where $C$ is the censoring time, and $\delta = I(T<C)$. Let $\{(X_i,\delta_i,Z_i),i=1,\cdots,n\}$ be $n$ independent realizations of $(X,\delta,Z)$.

In the presence of censoring, event status at $t_0$, $I(T_i<t_0)$, may not be observed for some subjects. We suggest using the inverse probability weighting (IPW) \cite{uno2007evaluating,lawless2010estimation} to account for censoring. The time-dependent PPV and TPF are estimated by
$$
\ppvhat_{t_0}(z) = \frac{\sum_{i=1}^n \what_{t_0,i} I(Z_i \geq z)I(X_i < t_0)}{\sum_{i=1}^n I(Z_i\geq z)}\quad \hbox{and}\quad \tpfhat_{t_0}(z) = \frac{\sum_{i=1}^n \what_{t_0,i} I(Z_i\geq z)I(X_i < t_0)}{\sum_{i=1}^n \what_{t_0,i}I(X_i < t_0)},
$$
where $\what_{t_0,i}$ is the inverse of the estimated probability that the time-dependent event status $I(T_i < t_0)$ is observed, specifically
$$
\what_{t_0,i} = \frac{I(X_i < t_0)\delta_i}{\Gcalhat(X_i)} + \frac{I(X_i \geq t_0)}{\Gcalhat(t_0)},
$$
where $\Gcalhat(c)$ is a consistent estimator of the survival function of the censoring time, $\Gcal(c) = Pr(C \geq c)$. Note that the proposed estimator does not imposes any assumptions on the relationship between the risk score $Z$ and the event time $T$. Under the assumption of independent censoring, i.e., the censoring time $C$ is independent of both the event time $T$ and the risk score $Z$, $\Gcal(c)$ can be obtained by the nonparametric Nelson-Aalen or Kaplan-Meier estimator. If the censoring time $C$ depends on the risk score $Z$, additional model assumptions might be required. For example, a proportional hazards (PH)  model could be fit to estimate $\Gcal_z(t) = Pr(C \geq c \mid Z=z)$.

Based on the estimated $\ppv_{t_0}(z)$ and $\tpf_{t_0}(z)$, $\ap_{t_0}$ can be estimated by
\begin{equation}\label{ap-est}
\widehat{\ap}_{t_0} = \frac{\sum_{j=1}^nI(X_j \leq t_0)\hat{w}_{t_0,j} \ppvhat_{t_0}(Z_j)}{\sum_{j=1}^nI(X_j \leq t_0)\hat{w}_{t_0,j}}.
\end{equation}
\cite{uno2007evaluating} shows that $\ppvhat_{t_0}(z)$ and $\tpfhat_{t_0}(z)$ are both consistent estimators, and thus,  $\widehat{\ap}_{t_0}$ is also a consistent estimator of $\ap_{t_0}$ for any given value of $t_0$.

In practice, we often deal with discrete risk scores, where tied risk scores are common. To accommodate risk scores with ties, following \cite{pepe2003statistical}, we modify the above estimator (\ref{ap-est}) by replacing $\ppvhat_{t_0}(Z_j)$ with
$$
\ppvtilde_{t_0}(Z_j) = \frac{\sum_{i=1}^n \what_{t_0,i} \left\{I(Z_i > Z_j) + \frac{1}{2}I(Z_i=Z_j)\right\}I(X_i < t_0)}{\sum_{i=1}^n \left\{I(Z_i > Z_j) + \frac{1}{2}I(Z_i=Z_j)\right\}}.
$$
To construct confidence intervals, we suggest the nonparametric bootstrap \cite{efron1979bootstrap} method. Specifically, let $\widehat{\mbox{AP}}^{\mathbbm B}_{t_0} =\left\{\widehat{\mbox{AP}}^b_{t_0}, b=1,2,\cdots,B\right\}$ denote the estimated $\ap_{t_0}$ obtained from $B$ bootstrape resamples. A 95\% confidence interval (CI) for the $\ap_{t_0}$ is given as $(\widehat{\mbox{AP}}_{t_0}^{\mathbbm B,0.025}, \widehat{\mbox{AP}}_{t_0}^{\mathbbm B,0.975})$, where $\widehat{\mbox{AP}}_{t_0}^{\mathbbm B,0.025}$ and $\widehat{\mbox{AP}}_{t_0}^{\mathbbm B,0.975}$ are the 2.5\% and 97.5\% empirical percentiles of the $\widehat{\mbox{AP}}^{\mathbbm B}_{t_0}$, respectively.

\subsection{Comparing two risk scores}
We consider comparing two risk scores $Z_1$ and $Z_2$ on the $\ap_{t_0}$ scale. In many studies, both risk scores $Z_1$ and $Z_2$ are calculated for each individual. With the paired data, we can quantify the relative predictive performance of $Z_1$ vs. $Z_2$, using the ratio of their respective time-dependent AP,
 $$\text{rAP}_{t_0}=\ap_{Z_1,t_0}/\ap_{Z_2,t_0},$$
 where $\ap_{Z_1,t_0}$ and $\ap_{Z_2,t_0}$ denote the time-dependent AP for $Z_1$ and $Z_2$ at $t_0$ respectively. For a single risk score $Z$, the ratio $AP_{Z,t_0}/\pi_{t_0}$ can be regarded as the relative predictive performance of $Z$ compared to a non-informative risk score.

The AP ratio $\text{rAP}_{t_0}$ can be estimated by $\widehat{\text{rAP}}_{t_0}=\widehat{\ap}_{Z_1,t_0}/\widehat{\ap}_{Z_2,t_0}$, where $\widehat{\ap}_{Z_1,t_0}$ and $\widehat{\ap}_{Z_2,t_0}$ are the the nonparametric estimator $\widehat{\ap}_{t_0}$ in (\ref{ap-est}) of $Z_1$ and $Z_2$ respectively.  The standard percentile bootstrap method descried in Section \ref{sec:single} can be used to construct a CI for $\text{rAP}_{t_0}$ or test $H_0: \text{rAP}_{t_0}=1$ for any given time point $t_0$. Specifically, the CI could be obtained based on the empirical percentiles of the $B$ bootstrap counterparts of of $\widehat{\text{rAP}}_{t_0}$, denoted by $\widehat{\text{rAP}}^b_{t_0}=\widehat{\ap}_{Z_1,t_0}^b/\widehat{\ap}_{Z_2,t_0}^b$, where $\widehat{\ap}_{Z_1,t_0}^b$  and $\widehat{\ap}_{Z_2,t_0}^b$ are the estimated $\ap_{t_0}$ for $Z_1$ and $Z_2$ based on the same bootstrap resample, $b=1,\cdots,B$.

\section{Simulation study} \label{sec:simulation}

\begin{figure}[h]
\centering
\includegraphics[width=5.0in]{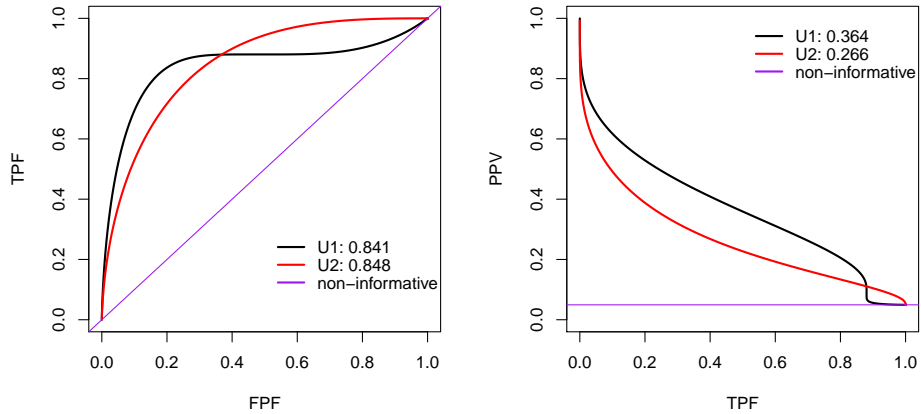}
\caption{The ROC curves in the left panel and the precision-recall curves in the right panel for the two risk scores $U_1$ and $U_2$ at $t_0=8$ when the event rate is 5\%. The numbers shown in graph correspond to the AUC and the AP values.}\label{fig:1}
\end{figure}

We conducted a simulation study to examine the performance of the time-dependent AP estimator in finite samples. In this simulation study, we considered two risk scores $U_{1}$ and $U_{2}$. They were generated from a standard normal distribution $N(0, 1)$. The event time associated with both risk scores for the ith subject was generated from the following model
$$\log(T_i) = 7.2 - 1.1 U_{i1} - 2.5 U_{i2} - 1.5 log(U_{i1}^2) + \epsilon_T,$$
where $\epsilon_T \sim  N(0, 1.5)$. This setting provides an example where the ROC curves of the two risk scores cross at time $t_0=8$, shown in Figure~\ref{fig:1}, with $\auc_{U_1, t_0}$ and $\auc_{U_2, t_0}$ are similar in values. On the other hand, the PR curve of $U_1$ dominates that of $U_2$ over the most range of the TPR with $\ap_{t_0}$ of $U_1$ greater than that of $U_2$.

The censoring time $C_i$  was generated following $C_i = \min (A_i, B_i+1)$ where, $A_i \sim  Uniform(0, 50)$, and $B_i \sim  Gamma(25,0.75)$. This configuration results in about 50\% of censoring overall. Let $X_i = \min(T_i, C_i)$, $\delta_i = I(T_i \leq C_i)$. In this setting, the censoring time is independent of both the event time and risk scores.

We considered three prediction time points $t_0$ where the corresponding event rates are, $r=P(T_{i} < t_{0})$,  0.01, 0.05 and 0.1, respectively. To allow a reasonable number of events by $t_0$, we generated the data $\{(X_i, \delta_i, U_{1i}, U_{2i}), i = 1,...,n\}$ with sample size $n$ being 2000 and 5000 (Tables~\ref{table:1} and \ref{table:2}). In each table, we report the following summary statistics based on 1000 repetitions: bias, empirical standard error (ESE) of the estimator, average standard errors from bootstrap ($ASE^b$), and the empirical coverage probability ($ECOVP^b$) of 95\% confidence intervals obtained from 1000 bootstrap resamples as described in Section~\ref{sec:estimator}.

These results show that the estimators of both time-dependent AP and ratio of AP, $AP_1/AP_2$, between the two risk scores had small biases in all $t_0$ values and different sample sizes. The bias decreases with increasing event rate and increasing sample size. Also, the standard errors $ASE^b$ obtained from bootstrap were close to the empirical standard errors. Thus, the confidence intervals attained the nominal coverage probabilities for both smaller sample size 2000 and larger sample size 5000.

We remark that this simulation provides an illustrative example of the relationship between ROC curve and PR curve as well as the relationship between the AUC and the AP investigated in \cite{davis2006}. When the ROC curves of two competing risk scores cross, the PR curves cross too. In situations like this, the AUC and the AP may rank the risk scores differently. In our simulation setting,  $U_2$ outperforms $U_1$ according to the AUC, which indicates that $U_2$ is better at discriminating between cases and controls. On the other hand, $U_1$ outperforms $U_2$ according to the AP, which suggests that $U_1$ is a better screening tool for stratifying subjects into different risk groups. 

\begin{table}[h]
\centering
\caption{Results of simulation with sample size 2000.}\label{table:1}

{\tabcolsep=4.25pt
\begin{tabular}{cm{1in}<{\centering}ccccccc}

\hline
$t_0$ & Event rate  & & TRUE & BIAS & ESE & $ASE^b$ & $ECOVP^b(\%)$ \\
\hline
\hline

 0.5 & 0.0101
 & $\ap_1$ & 0.182 & 0.0365 & 0.0810 & 0.0795 & 92.3 \\
\cline{3-8}
 & & $\ap_2$ & 0.124 & 0.0339 & 0.0689 & 0.0678 & 93.0	 \\
\cline{3-8}
 & & $\text{rAP}$ & 1.47 & 0.4890 & 1.5300 & 1.7600 & 95.1 \\
\hline
8 & 0.0495
 & $\ap_1$ & 0.364 & 0.0096 & 0.0527 & 0.0516 & 92.5 \\
\cline{3-8}
 & & $\ap_2$ & 0.266 & 0.0129 & 0.0452 & 0.0450 & 93.4	 \\
\cline{3-8}
 & & $\text{rAP}$ & 1.37 & 0.0140 & 0.3290 & 0.3320 & 95.7 \\
\hline
36 & 0.0991
 & $\ap_1$ & 0.462 & 0.0098 & 0.0534 & 0.0558 & 95.9  \\
\cline{3-8}
 & & $\ap_2$ & 0.375 & 0.0118 & 0.0493 & 0.0501 & 94.5 \\
\cline{3-8}
 & & $\text{rAP}$ & 1.23 & 0.0135 & 0.2310 & 0.2420 & 94.9 \\
\hline
\end{tabular}}
\end{table}

\begin{table}[h]
\centering
\caption{Results of simulation with sample size 5000.}\label{table:2}

{\tabcolsep=4.25pt
\begin{tabular}{cm{1in}<{\centering}ccccccc}

\hline
$t_0$ & Event rate  & & TRUE & BIAS & ESE & $ASE^b$ & $ECOVP^b(\%)$ \\
\hline
\hline

 0.5 & 0.0101
 & $\ap_1$ & 0.182 & 0.0185 & 0.0500 & 0.0504 & 93.1 \\
\cline{3-8}
 & & $\ap_2$ & 0.124 & 0.0155 & 0.0416 & 0.0417 & 94.8	 \\
\cline{3-8}
 & & $\text{rAP}$ & 1.47 & 0.1550 & 0.7060 & 0.7600 & 93.8 \\
\hline
8 & 0.0495
 & $\ap_1$ & 0.364 & 0.0042 & 0.0337 & 0.0333 & 92.9 \\
\cline{3-8}
 & & $\ap_2$ & 0.266 & 0.0049 & 0.0291 & 0.0288 & 93.7	 \\
\cline{3-8}
 & & $\text{rAP}$ & 1.37 & 0.0060 & 0.2160 & 0.2100 & 95.4 \\
\hline
36 & 0.0991
 & $\ap_1$ & 0.462 & 0.0034 & 0.0354 & 0.0346 & 95.5 \\
\cline{3-8}
 & & $\ap_2$ & 0.375 & 0.0037 & 0.0310 & 0.0313 & 94.1 \\
\cline{3-8}
 & & $\text{rAP}$ & 1.23 & 0.0051 & 0.1490 & 0.1510 & 95.0 \\
\hline
\end{tabular}}
\end{table}

%


\section{Data Analysis} \label{sec:data-analysis}
In this section, we illustrate the use of AP$_{t_0}$ metric with a data set from the Childhood Cancer Survivor Study \cite{robison2009childhood}. This cohort follows children who were initially treated for cancer at 26 US and Canada institutions between 1970 and 1986 and who survived at least 5 years after their cancer diagnosis. Among the survivors, cardiovascular disease has been recognized as a leading contributor to morbidity and mortality \cite{oeffinger2006chronic}. To inform future screening and intervention strategy for congestive heart failure (CHF) in this population, \cite{chow2015individual} developed several risk score systems using the CCSS data  and validated them on external cohorts. For the purpose of illustration, we chose two of these risk scores and evaluated their predictive performance using the proposed AP$_{t_0}$.

We included 11,457 subjects  in our analysis from the CCSS study who met the original study inclusion criteria and had both risk scores. In this data, a total of 248 subjects experienced the CHF. Between the two risk scoring systems we focused on in this data analysis, the simpler model used information on age at cancer diagnosis, sex,  whether the patient was exposed to chest radiotherapy, and whether the patient was exposed to a particular chemotherapy agent. We refer to this model as the simple model. The more elaborate model, known as the heart dose model, included detailed clinical information on the average radiation dose to the heart and the cumulative dose of the specific chemotherapy agent, along with age at diagnosis and sex. This is an example where a simple risk score system utilizes minimum treatment information and can be used for any patient by virtually all clinicians, while the more complex risk score system demands specific dose information which may not be readily available to clinicians providing long-term follow-up care. We used the original risk scores of the simple model and the heart dose model from the reference study \cite{chow2015individual}. Briefly, these scores were constructed via linear combinations of the corresponding covariates, where the regression coefficients were obtained from Poisson regression models.

\begin{figure}[!h]
\begin{center}
\includegraphics[width=5.0in]{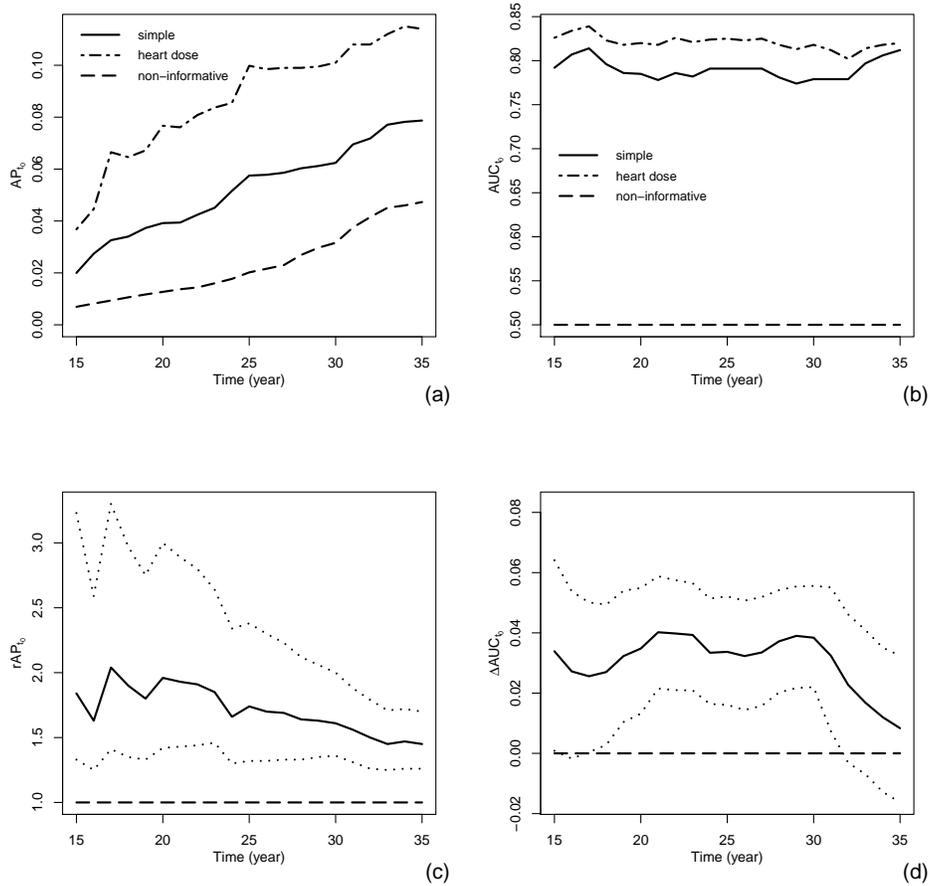}
\caption{CCSS Data analysis: panel (a) shows the estimates of the time-dependent AP for the simple model $\ap_{s,t_0}$ and heart dose model $\ap_{h,t_0}$; panel (b) shows the estimates of the time-dependent AUC for the simple model $\auc_{s,t_0}$ and heart dose model $\auc_{h,t_0}$; panel (c) shows the estimates of $\text{rAP}_{t_0}=\ap_{h,t_0}/\ap_{s,t_0}$, the ratio of the time-dependent AP of the heart dose model over that of the simple model; panel (d) shows the estimates of $\Delta \auc_{t_0}=\auc_{h,t_0}-\auc_{s,t_0}$, the difference of the time-dependent AUC between the heart dose model and the simple model. The dotted lines in panels (c) and (d) represent the pointwise 95\%CI for $\text{rAP}_{t_0}$ and $\Delta \auc_{t_0}$, respectively.} \label{fig:CCSS}
\end{center}
\end{figure}

Table~\ref{tab:CCSS} reports the estimated AP$_{t_0}$ with 95\% CIs for both the simple model (denoted by $\ap_{s,t_0}$) and heart dose model (denoted by $\ap_{h,t_0}$) at $t_0$ = 20 and 35 years post-diagnosis where the corresponding estimated event rates were 1.3\%  and 4.7\% respectively. These two models were compared using the ratio of AP, i.e. $\text{rAP}_{t_0} = \ap_{h,t_0}/\ap_{s,t_0}$. In addition, we also provided the estimated time-dependent AUC ($\auc_{t_0}$) at these two time points as well as the difference of AUCs between these two models $\Delta\auc_{t_0}=\auc_{h, t_0} - \auc_{s, t_0}$. To illustrate the time-varying performance for each model as well as the comparison between these two models over time, $\ap_{t_0}$, $\auc_{t_0}$, $\text{rAP}_{t_0}$ and  $\Delta\auc_{t_0}$ versus $t_0$ were plotted in Figure~\ref{fig:CCSS}. Note that we assumed independent censoring in estimating the AP and the AUC.

The results in Table~\ref{tab:CCSS} show that the heart dose model outperforms the simple model at both time points. For example, the estimated $\ap_{20}$ of the heart dose model is 0.075, which indicates that by 20 years post-diagnosis, using the risk score from the heart dose model, we expect that on average 7.5\% subjects with a high risk score (compared to the risk score of a randomly selected case) will experience heart failure. This AP is almost six times of the event rate 1.3\%, which corresponds to the AP of a non-informative risk score system. In contrast, the estimated $\ap_{20}$ for the simple model is 0.039, roughly half of that of the heart dose model ($r\ap_{20}$=1.96, 95\%CI:1.41-2.88). At 35 years post diagnosis, the heart dose model is significantly better than the simple model with $r\ap_{35}$ = 1.45 (95\%CI: 1.26 - 1.70). Indeed, The plots (c) and (d) in Figure \ref{fig:CCSS} show that in terms of the $\ap_{t_0}$, the heart dose model outperforms the simple model at identifying the high risk subjects from the targeted population at all time points considered. On the other hand, the AUCs are similar between these two models, as seen in plot (d). Especially, $\Delta\auc$ is not significantly different from 0 at the beginning and towards the end of the time period that were considered,  For example $\Delta\auc_{35} = 0.01$ (95\% CI: -0.02 - 0.03, p=0.47), shown in Table \ref{tab:CCSS}. It suggests that according to the AUC, there is not much performance difference between the two models. If, due to incorporating more information, the heart dose model is indeed superior to the simple model in terms of identifying the high risk individuals, the results in Table~\ref{tab:CCSS} and Figure \ref{fig:CCSS} implies that the AP is a better metric for discriminates the risk prediction performance than the AUC does.

\begin{table}[h]
\centering
\caption{Estimated $\ap_{t_0}$ and AUC$_{t_0}$ with 95\% CIs for two risk scoring systems at $t_0$ = 20 and 35 years, respectively. The comparison of APs is measured by $\text{rAP}$ and the comparison of AUCs is measured by  $\Delta\auc$. }\label{tab:CCSS}

{\tabcolsep=4.25pt
\begin{tabular}{ccccc}
\hline
$t_0$ & Event rate & Risk score system  & AP & AUC \\
\hline
\hline

 20 years &  0.013
 & Simple &  0.039 (0.029, 0.052) & 0.79 (0.75, 0.82) \\
\cline{3-5}
 & &  Heart dose &  0.077 (0.050, 0.127) & 0.82 (0.78, 0.86)\\
\cline{3-5}
 & &  Comparison &  1.96 (1.41, 2.88) & 0.03 (0.01, 0.05)\\
\hline
35 years & 0.047
 & Simple & 0.079 (0.066, 0.095) & 0.81 (0.78, 0.84) \\
\cline{3-5}
 & &   Heart dose &  0.114 (0.092, 0.144) & 0.82 (0.78, 0.85) \\
\cline{3-5}
 & &  Comparison &  1.45 (1.26, 1.70) & 0.01 (-0.02, 0.03) \\
\hline

\end{tabular}}
\end{table}

\section{Discussion} \label{sec:discussion}
One of the main goals of clinical risk prediction is to screen the asymptomatic population and to stratify them for tailored intervention. \textit{Prospective} accuracy measures such as PPV$_{t_0}$ is preferred for this purpose. However, the calculation of PPV$_{t_0}$ demands a threshold for continuous risk scores, which can create practical difficulties for evaluating risk score systems, especially when more than two systems are compared. In this paper, we defined and interpreted $\ap_{t_0}$, which is the area under the time-dependent precision-recall curve, for event time data. We proposed a nonparametric estimator of $\ap_{t_0}$ and a ratio estimator of $\ap_{t_0}$ for comparing two competing risk score systems. We suggested the use of the bootstrap method for inference, which is broadly applicable in practical settings. We also developed an R package \texttt{APtools} for download available in CRAN which implements our method for binary and survival outcomes.

The AUC has been the most widely used performance metric in the clinical research community. A number of authors have pointed out that the AUC is informative on the classification performance and discrimination power \cite{gail2005criteria, zheng2008time, yuan2015threshold}, but not an appropriate metric for assessing the prospective accuracy performance \cite{moskowitz2004quantifying,zheng2008time}. Consistent with the criticism on the insensitivity of the AUC in evaluating risk prediction models (See \cite{cook2007use}), our data analysis illustrated that using the AUC as the metric, the performance of the simple model and the heart dose model appears close. However,based on the AP, the heart dose model outperforms the simple model and could be preferred in clinical screening. Thus the conclusion reached based on the AUC in this example may mislead researchers and clinicians.

It should be noted that when comparing different risk score systems, the ranking of their AUCs and APs are not necessarily concordant; our simulation study in Section 4 gives such an example. As the purpose of this article is to introduce a time-dependent $\ap_{t_0}$, we refer readers to \cite{davis2006} for insight on the relationship between the ROC curve and PR curve and to \cite{su2015relationship} where the relationship between the AUC and the AP is illuminated. Thus, risk scores which perform well in separating cases from controls may perform poorly in identifying a higher risk subpopulation, which is the goal of screening. Compared to the AUC, the AP as the summary metric of PPV is better suited in evaluating the usefulness of the risk scores and comparing the \textit{prospective }prediction performance among competing risk scores, when the objective is \textit{screening} through risk stratification.

\cite{zheng2008time, zheng2010semiparametric} proposed to use the curves of PPV$_{t_0}$ versus risk score quantiles as an assessment tool for quantifying prospective prediction accuracy. One curve corresponds to one particular value of $t_0$, which limits its ability to assess the accuracy across time points.  In contrast, plotting AP$_{t_0}$ against time could facilitate visualizing the performance of different risk score systems over time in one single plot.

Unlike the AUC, the AP is event rate dependent and should be estimated in a prospective cohort or population-based study. AP cannot be estimated from a case-control study; the estimate will be of very little use because the prevalence rate is artificially fixed by the study design. While the range of the AUC is always between 0.5 and 1, the range of AP is between the event rate and 1. While AP's wide range could be advantageous in differentiating risk score systems, caution is needed when re-evaluating risk score systems in other study populations for the same outcome. This is because the underlying event rate may differ among populations. Thus, it is possible that AP will select different risk score systems as superior for the same outcome in different study populations.

For future work, we will consider estimating the time-dependent AP with multiple markers, as well as the incremental value of AP by adding new markers on top of an existing risk profile.
In addition, similar to the partial AUC,  partial AP could be defined as the area over a certain range of interest, such as those at the low values of TPF where PPV is typically high. Competing risk in the context of AP is another topic that needs to be addressed for event time data.

\section*{Acknowledgments}
The authors would like to thank Yan Chen for data support. Dr. Zhou's research is supported by the Natural Sciences and Engineering Research Council of Canada. Dr. Yuan's research is supported by the Canadian Institutes of Health Research. The CCSS was supported by the National Cancer Institute (CA55727, G.T. Armstrong, Principal Investigator).

\bibliographystyle{wileyj}
\bibliography{AP}

\end{document}